\documentclass[journal]{IEEEtran}
\usepackage[cmex10]{amsmath}
\usepackage{graphicx}
\usepackage{cite}
\usepackage{epstopdf}
\usepackage{algorithm}
\usepackage{algorithmic}
\usepackage[caption = false]{subfig}
\usepackage{booktabs}
\usepackage{amsfonts}
\usepackage{color}
\newtheorem{lem}{\textbf{Lemma}}
\newtheorem{theorem}{\textbf{Theorem}}
\newtheorem{proposition}{\textbf{Proposition}}

\begin{document}
\title{Energy Efficiency Analysis of Heterogeneous Cellular Networks with Extra Cell Range Expansion}
\author{Yuhao~Zhang, Zhiyan~Cui, Qimei~Cui,~\IEEEmembership{Senior~Member,~IEEE,} \\
Xinlei~Yu, Yinjun~Liu, Weiliang~Xie, and Yong~Zhao
\IEEEcompsocitemizethanks{
\IEEEcompsocthanksitem The work was supported in part by National Nature Science Foundation of China Project (Grant No.61471058), Key National Science Foundation of China (61231009), Hong Kong, Macao and Taiwan Science and Technology Cooperation Projects (2016YFE0122900) and the 111 Project of China (B16006)​.
\IEEEcompsocthanksitem Y. Zhang, Z. Cui, Q. Cui, X. Yu, and Y. Liu are with the School of Information and Communication Engineering, Beijing University of Posts and Telecommunications, Beijing 100876, China (e-mail: cuiqimei@bupt.edu.cn).
\IEEEcompsocthanksitem W. Xie and Y. Zhao are with the China Telecom Corporation Limited, Beijing 100140, China (e-mail: \{xiewl.bri, zhaoyong.bri\}@chinatelecom.cn).}}

\maketitle
\begin{abstract}
The split control and user plane is key to the future heterogeneous cellular network~(HCN), where the small cells are dedicated for the most data transmission while the macro cells are mainly responsible for the control signaling. Adapting to this technology, we propose a general and tractable framework of extra cell range expansion~(CRE) by introducing an additional bias factor to enlarge the range of small cells flexibly for the extra offloaded macro users in a two-tier HCN, where the macro and small cell users have different required data rates. Using stochastic geometry, we analyze the energy efficiency~(EE) of the extra CRE with joint low power transmission and resource partitioning, where the coverages of EE and data rate are formulated theoretically. Numerical simulations verify that the proposed extra CRE can improve the EE performance of HCN, and also show that deploying more small cells can provide benefits for EE coverage, but the EE improvement becomes saturated if the small cell density exceeds a threshold. Instead of establishing the detail configuration, our work can provide some valuable insights and guidelines to the practical design of future networks, especially for the traffic offloading in HCN.
\end{abstract}

\begin{IEEEkeywords}
Energy efficiency, heterogeneous network, extra cell range expansion, interference management, stochastic geometry.
\end{IEEEkeywords}

\section{Introduction}\label{sec:1}
The mobile wireless communications, which require large system capacity, strong signal coverage, and high energy efficiency (EE), have developed swiftly and experienced explosive data growth over the past decade~\cite{ref1,ref16}. Driven by this trend, satisfying the improved quality of service~(QoS), the cellular networks now have to deliver numerous data information to the users with higher spectral efficiency~(SE). Therefore, the architecture of multi-tier heterogeneous cellular network~(HCN), based on small cells, has been proposed as a promising technology to increase the network capacity and avoid coverage holes~\cite{ref4}. Moreover, due to the rapidly growing energy consumption in wireless networks, the research of EE has attracted wide attention all over the world~\cite{ref2,ref3}. It is known that HCN with small cells can solve the physical scarcity of radio frequency and increase the system capacity to some extent, but at the cost of deploying more hardware infrastructures, which leads to higher energy consumption consequently~\cite{ref9,ref17}. As a result, due to the trade-off between the benefit and cost, the EE of HCN is quite interesting and worthy of research, which is mainly discussed in this paper.

In HCN, the densely deployed small cells are usually lightly loaded because of their low transmit power compared with macro base stations~(BSs). In order to take full advantage of small cells, a biasing technique known as cell range expansion~(CRE) can effectively offload more traffic to small cells to get better system capacity~\cite{ref5}. In the future wireless networks, the control and user planes are decoupled, where the small cells with small and strong coverage are dedicated for the most data transmission while the macro cells with wide and weak coverage are responsible for the control signaling and the remaining data information~\cite{ref18}. Therefore, more users, including some original macro users, should be served by small cells, which can provide higher achievable data rate due to the shorter transmission distance and the adequate resources. Considering this issue, the small cells need the wider and more flexible CRE to access more users of different types, but the traditional CRE cannot work efficiently anymore in this situation because directly increasing the bias factor cannot represent the data rate difference between small cells and macro cells. Therefore, the extra CRE with better utilization of small cells, which can provide more flexibility by introducing an additional bias factor, is proposed in this paper.

Compared with traditional CRE, the users in the extra expanded area will suffer more severe degraded signal-to-interference and noise ratio~(SINR) in co-channel deployments since they are located much closer to the macro BSs and thus receive stronger interference power. Therefore, the suitable interference management strategies are required urgently to overcome this inter-tier interference problem. One typical and fundamental solution is to plan and design the spectrum usage for different cells in the networks, such as resource partitioning where the macro cells are prevented from transmitting on certain fraction of resources with full power~\cite{ref19}. Therefore, in this protected fraction of resources, the offloaded users can be scheduled by small cells under much lower interference from the macro BSs, which can either stop data transmission or simply reduce the transmit power~\cite{ref8}.
\subsection{Related Work}\label{sec:1a}
Recently, the research of CRE in HCN has attracted more interests all over the world. In~\cite{ref14}, a general and tractable framework to model and analyze joint resource partitioning and offloading in two-tier cellular networks is proposed and the downlink rate distribution is derived, which reveal that the resource partitioning is necessary for offloading to improve the system performance. Adopting resource partitioning, an analytical framework with joint user association and sleep-mode BSs in HCN is proposed in~\cite{ref7}. Numerical results show that the proposed scheme can reduce the network energy consumption and improve the SINR coverage in co-channel HCN. In~\cite{ref10}, the resource fraction is optimized to minimize the inter-tier interference and maximize the EE of the networks, respectively. Moreover, in~\cite{ref6}, the theoretical expressions provide insights into the SINR coverage improvement by jointly applying transmit power reduction and resource partitioning strategies in HCN.

The EE research of CRE is relatively less in the existing literatures, particularly on the average EE coverage analysis with stochastic geometry. The energy efficient user association scheme~\cite{ref24} and resource allocation scheme~\cite{ref25} are proposed to maximize the overall EE of HCN by designing effective iterative algorithms. In~\cite{ref23}, the EE analysis is carried out in the cache-enabled HCN with the control and user plane split, and the expressions of coverage probability, throughput and EE are derived analytically as the functions of the cache ability, search radius and backhaul limitation. In~\cite{ref26}, the total EE is evaluated in cellular-WLAN heterogeneous network employing CRE technique, and with EE optimization, it shows that the trade-off between the fairness of user throughput and the system EE can achieve better balance by optimizing the bias factor of each access point individually.

However, only the traditional CRE is considered and investigated in the above works, which also pay relatively less attention to the decoupled control and user plane, and do not distinguish the data rate targets for different tiers in HCN as well. To the best of our knowledge, the extra CRE based on the traditional CRE with joint low power transmission and resource partitioning in HCN has never been studied, not to mention the EE analysis using stochastic geometry, which will be proposed and discussed in this paper.
\subsection{Contributions}\label{sec:1b}
In this paper, considering the decoupled control and user plane in the future wireless networks, where the users connecting to macro BSs and small cells have different required data rates, we propose the general extra CRE with joint low power transmission and resource partitioning in HCN. By introducing another association bias factor for further user offloading to enlarge the small cell range, the size of the extra expanded range is controlled flexibly by the two bias factors jointly, which can provide better system performance consequently, as shown in the simulations. This paper is focused on the EE analysis of HCN with extra CRE, and using stochastic geometry we formulate the average EE coverage of the system by deriving the SINR and data rate coverages. Rather than establishing detail configuration of CRE, our work provides valuable insights and guidelines to the design of the future wireless networks, especially for traffic offloading in HCN.

The contributions of this paper are summarized as follows.
\begin{itemize}
  \item we propose a general framework of extra CRE, containing an additional bias factor, with joint transmit power reduction and resource partitioning in the two-tier HCN, which can accommodate users with different required data rates and can be adopted for the decoupled control and user planes in future networks flexibly.
  \item with stochastic geometry, we analyze the EE of HCN with extra CRE and formulate the theoretical average EE coverage as well as the SINR and data rate coverages, which are key to evaluate the system performance.
  \item we show that the proposed extra CRE can improve the system performance of HCN, including the EE coverage, and deploying more small cells can increase the EE coverage as well but the EE improvement becomes saturated if the small cell density exceeds a threshold.
\end{itemize}
\subsection{Organization and Notations}\label{sec:1c}
In the rest of the paper, Section~\ref{sec:2} presents the system model and Section~\ref{sec:3} describes our proposed extra CRE with joint low power transmission and resource partitioning. The coverages of SINR, data rate and EE are formulated in Section~\ref{sec:4}. Simulation results are provided in Section~\ref{sec:5}, followed by the conclusion and future work in Section~\ref{sec:6}.

In this paper, $\mathbb{I}\left( x \right)$ denotes the indicator of event $x$, i.e., if the event $x$ is true, $\mathbb{I}\left( x \right)$ equals to 1, otherwise $\mathbb{I}\left( x \right)$ equals to 0. And $\mathbb{P}\left( x \right)$ is the probability of event $x$. $\mathbb{E}\left( \cdot \right)$ is the mean function and ${\cal L}_I \left( \cdot \right)$ is the Laplace transform of interference $I$. Furthermore, we denote the empty set by $\emptyset$.

\section{System Model}\label{sec:2}
\begin{figure}[!t]
\centering
\includegraphics[scale=0.28]{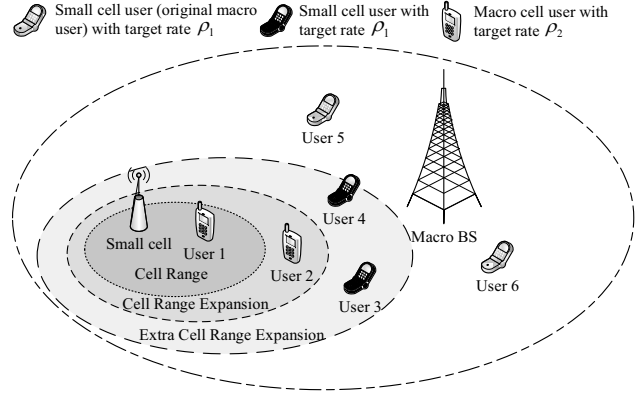}
%\vspace{-0.5cm}
\caption{The system model of extra cell range expansion.}\label{fig.systemmodel}
\vspace{-3mm}
\end{figure}

In this paper, we consider a downlink heterogeneous wireless network, which contains two independent network tiers of BSs, i.e., macro cells and small cells, with different deployment densities and transmit powers. Without loss of generality, it is assumed that the macro cells constitute tier 1 and the small cells constitute tier 2. The spatial distribution of BSs of tier $k\left( {k = 1,2} \right)$ in this network conforms a two-dimensional homogeneous Poisson point processes~(PPP) ${\Phi _k}$ with the density of ${\lambda _k}$, and the users, denoted by $u$, are located in the researched area according to another independent homogeneous PPP with the constant density of ${\lambda _u}$. All the BSs within the tier $k$ have the same transmit power, denoted by ${{{P_k}}\left({k = 1,2}\right)}$. In this paper, we study the typical user located at the origin $(0,0) \in \mathbb{R}^2$ for average performance of the network, which is allowed by Slivnyak's theorem~\cite{ref15}.

As shown in~Fig.~\ref{fig.systemmodel}, with traditional cell range expansion~\cite{ref14,ref8,ref19}, the users connect to BSs according to the maximum biased received power~(BRP), and then can be decomposed into macro users (like users 1 and 2 in Fig.~\ref{fig.systemmodel}) and small cell users (like users 4 and 5 in Fig.~\ref{fig.systemmodel}). Since the initial purpose of deploying small cells in practice is to provide high data rate coverage, the data rate requirement of the small cell users is generally larger than that of the users in macro cells~\cite{ref7}. Therefore, the required data rate for macro cell users can be denoted by ${\rho _1}$, and ${\rho _2}$ for small cell users, where ${\rho _1} \le {\rho _2}$. By proposing the extra CRE, the coverage of small cells can be divided into three disjoint areas including an extra expanded area, where the users (like users 3 and 4 in Fig.~\ref{fig.systemmodel}) that should have been served by macro cells will be offloaded to the small cells.
\subsection{Channel Model}\label{sec:2a}
The wireless channel from a typical user to its serving BS with the transmission distance of $x$ is modeled by the standard large scale path loss and Rayleigh fading. The path loss factor, denoted by ${\alpha _k}$, is different in each tier, and the independent complex Rayleigh fading coefficient, denoted by $g_x$, has zero mean and unit variance. For simplicity, the newly defined coefficient ${h_x}={{\left| {{g_x}} \right|}^2}$ is i.i.d exponentially distributed with unit mean, denoted by ${h_{{x}}}\sim\exp{(1)}$. Therefore, a typical user located at the origin receives the power of ${P_k}{h_x}{x^{ - \alpha_k }}$ from its serving BS. $W$ is the system bandwidth, and $n$ is the additive white Gaussian noise~(AWGN) with power ${\sigma ^2}=N_0 W$, where $N_0$ is the power spectral density~(PSD) of the noise.
\subsection{Power Consumption Model}\label{sec:2b}
We adopt the power consumption model used in~\cite{ref12}, where the power consumption for each BS is fixed, which can be interpreted as a constant average BS power consumption since a large number of BSs are considered. With this power model, the power consumption of the $i$-th BS during downlink transmission can be given by
\begin{equation}\label{eq1}
{P_{i,{\rm total}}} = {a_i}{P_i} + {b_i},
\end{equation}
where coefficient ${a_i}$ accounts for the power consumption that scales linearly with the transmit power, whereas ${b_i}$ represents the static power consumed by signal processing, battery backup, and site cooling. It is noted that this power consumption model reflects the fact that the average power consumption of a BS comprises both transmit and static powers.
\section{Extra Cell Range Expansion}\label{sec:3}
In this paper, to improve the system performance and adapt to the trend of the split control and user plane, we consider offloading more extra macro users to be served by small cells, i.e., some macro users with lower target data rate ${\rho _1}$ will connect to small cell BSs.
\subsection{User Association}\label{sec:3a}
The user association is based on the biased received power, i.e., the user will connect to the BS providing the maximum biased received power, where the bias factor is used to be multiplied by the received signal strength during the cell selection. In this paper, the accomplishment of the extra cell range expansion is fulfilled in two steps: the traditional biased user association is first achieved~\cite{ref14,ref8,ref19} and then the macro users in the extra expanded cell range will be offloaded to the small cells by introducing another association bias factor.

The bias factor ${B_2}$ is used for the traditional biased user association, i.e., the first phase of extra CRE, based on which, the user will be originally associated with the BS of tier $k$,
\begin{equation}\label{eq25}
k= {\rm arg} \max \{ {P_1}D_1^{ - \alpha_1 }, {P_2}{B_2}D_2^{ - \alpha_2 } \},
\end{equation}
where ${D_k}\left( {k = 1,2} \right)$ denotes the distance between the typical user and its nearest BS of tier $k$. Therefore, the target data rate ${{\rho}}$ for the user can be obtained consequently, as given by
\begin{equation}\label{eq2}
{\rho} = \left\{ {\begin{array}{*{20}{c}}
   {{\rho _1},} & {{\rm if}\ {P_1}D_1^{ - \alpha_1 } \ge {P_2}{B_2}D_2^{ - \alpha_2 }},  \\
   {{\rho _2},} & {{\rm if}\ {P_2}{B_2}D_2^{ - \alpha_2 } > {P_1}D_1^{ - \alpha_1 }}.
\end{array}} \right.
\end{equation}

Another association bias factor, denoted by ${B_1}$, is introduced here to offload extra macro users with target data rate ${\rho_1}$ to small cells; in other words, it can determine how many original macro users are still served by macro BSs and how many will be transferred to small cells. It is obvious that ${B_1} > {B_2}$ must be satisfied.

The non-extra biased small cell users (called original small cell users, i.e., users 1 and 2 in Fig.~\ref{fig.systemmodel}) with target data rate ${\rho_2}$ are collected into a user set, denoted by ${u_{\bar D}}$. Moreover, the set of users with target data rate ${\rho_1}$ is split into two disjoint sets, denoted by ${u_1}$ and ${u_D}$, where ${u_1}$ is the set of macro cell users (called macro users, i.e., users 5 and 6 in Fig.~\ref{fig.systemmodel}) and ${u_D}$ represents the set of users (called expanded small cell users, i.e., users 3 and 4 in Fig.~\ref{fig.systemmodel}) that are transferred from macro cells to small cells due to the extra cell range expansion. With the association scheme mentioned above, the mapping relationship between the user $u$ with data rate requirement ${\rho}$ and the sets ${u_1}$, ${u_D}$ and ${u_{\bar D}}$ is summarized, as given by
\begin{equation}\label{eq3}
u \hspace{-1mm} \in \hspace{-1mm}
\left\{ \begin{array}{ll}
\hspace{-2mm} {u_1}, & \hspace{-1mm} {\text{if}\ {\rho} \hspace{-1mm} = \hspace{-1mm} {\rho _1}\ \text{and}\ {P_1}D_1^{ - \alpha_1} \hspace{-1mm} \ge \hspace{-1mm} {P_2}{B_1}D_2^{ - \alpha_2 }},\\
\hspace{-2mm} {u_D}, & \hspace{-1mm} {\text{if}\ {\rho} \hspace{-1mm} = \hspace{-1mm} {\rho _1}\ \text{and}\ {P_2}{B_1}D_2^{ - \alpha_2} \hspace{-1mm} > \hspace{-1mm} {P_1}D_1^{ - \alpha_1} \hspace{-1mm} \ge \hspace{-1mm} {P_2}{B_2}D_2^{ - \alpha_2}},\\
\hspace{-2mm} {u_{\bar D}}, & \hspace{-1mm} {\text{if}\ {\rho} \hspace{-1mm} = \hspace{-1mm} {\rho _2}\ \text{and}\ {P_2}{B_2}D_2^{ - \alpha_2} \hspace{-1mm} > \hspace{-1mm} {P_1}D_1^{ - \alpha_1}},
\end{array} \right.
\end{equation}
where ${u_1} \cup {u_D} \cup {u_{\bar D}}$ is the set of all users. Furthermore, all the users associated with small cells are ${u_2} = {u_D} \cup {u_{\bar D}}$. It can be concluded that the extra CRE is determined by ${B_1}$ and ${B_2}$ jointly; in other words, the larger $\frac{B_1}{B_2}$ is, the wider the extra expanded range will be. Specifically, there is no extra CRE and the user set ${u_D}$ will be empty when ${B_1} = {B_2}$.
\subsection{Resource Partitioning and Transmit Power Reduction}\label{sec:3b}
\begin{figure}[!t]
\centering
\includegraphics[scale=0.48]{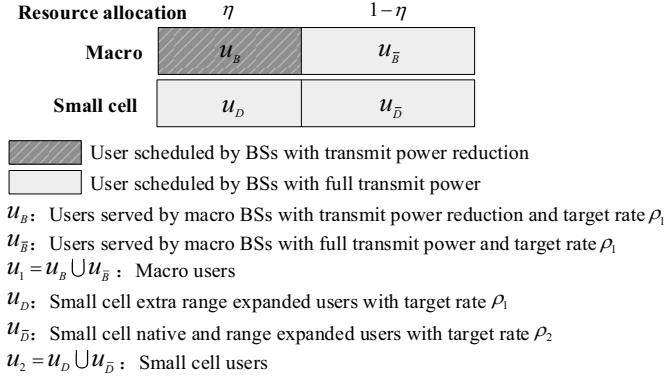}
\vspace{-6mm}
\caption{Resource allocation scheme.}\label{fig.1}
\vspace{-3mm}
\end{figure}

With extra CRE, the offloaded users (in user set ${u_D}$) will have further degraded SINR compared with their former situation, i.e., connecting to the macro BSs, since the useful signal becomes weaker and the original strong useful signal now contributes to the interference. To resolve this problem, we continue to combine the user association with the resource partitioning and transmit power reduction for better system performance~\cite{ref19,ref8}, as shown in Fig.~\ref{fig.1}. For clarity, the term ``resources'' used in this paper refers to a set of time/frequency 3GPP resource elements~\cite{ref20}. It is noted that the users in ${u_D}$, connecting to the small cells, are the exact users that need to be protected to ensure the desired gain of balancing load in HCN with extra CRE, and therefore, the transmit power reduction is only adopted in the macro BSs.

As for the resource allocation scheme in this paper, as shown in Fig.~\ref{fig.1}, the macro BSs share all the resources, and reduce the transmit power on $\eta\ (0 < \eta < 1)$ fraction resources to decrease the interference for the protected range extra expanded users, where the transmit power fraction $\beta\ (0 < \beta < 1)$ denotes the fraction of the full transmit power of tier 1 on the $\eta $ fraction of resources. For simplicity, ${u_B}$ is defined as the set of macro users scheduled in the $\eta $ fraction of resources with transmit power reduction, and ${u_{\bar B}}$ is the set of macro users scheduled in the $1 - \eta$ fraction of resources with full transmit power. In the small cell tier, $\eta$ is the fraction of resources allocated to the range extra expanded users, and the remaining $1 - \eta $ fraction of resources are shared by the non-extra expanded users. The scheduling type index $l \in \left\{ {B,\bar B,D,\bar D} \right\}$ is defined to represent the type of resources which a typical user will use. We also define a mapping relationship $M\left\{ {B,\bar B,D,\bar D} \right\} \to \left\{ {1,2} \right\}$ from the user set index to the serving tier index. Therefore, according to~\eqref{eq2}, the following relationships can be obtained as
\begin{equation}\nonumber
M\left( B \right) = M\left( {\bar B} \right) = 1,\ M\left( D \right) = M\left( {\bar D} \right) = 2.
\end{equation}

Considering the resource partitioning and transmit power reduction, the SINR of a typical user scheduled in the resource type $l$ can be formulated as
\begin{equation}\label{eq4}
\begin{aligned}
 {\rm SINR}_l & = \mathbb{I}\left( {l \in \left\{ B \right\}} \right)\frac{{\beta {P_1}{h_x}{x^{ - \alpha_1 }}}}{{\beta {I_{{r_1}}} + {I_{{r_2}}} + {\sigma ^2}}} \\
 & + \mathbb{I}\left( {l \in \left\{ {\bar B,\bar D} \right\}} \right)\frac{{{P_{M\left( l \right)}}{h_x}{x^{ - \alpha_{M\left( l \right)} }}}}{{{I_{{r_1}}} + {I_{{r_2}}} + {\sigma ^2}}} \\
 & + \mathbb{I}\left( {l \in \left\{ D \right\}} \right)\frac{{{P_2}{h_x}{x^{ - \alpha_2 }}}}{{\beta {I_{{r_1}}} + {I_{{r_2}}} + {\sigma ^2}}},
\end{aligned}
\end{equation}
where ${I_{{r_i}}}$ represents the cumulative interference from the tier $i$, which is presented by
\begin{equation}\nonumber
{I_{{r_i}}} = {P_i}\sum\limits_{r \in {\Phi _i}\backslash {b_0}} {{h_r}{r^{ - {\alpha _i}}}},
\end{equation}
where $b_0$ is the BS serving the typical user and $r$ is the distance between the BS and the typical user.
\section{Analysis of Energy Efficiency}\label{sec:4}
In this section, the EE performance of the network is characterized by using the criterion called energy efficiency coverage~(EEC)~\cite{ref13}, which is defined as
\begin{equation}\nonumber
C\left( \tau  \right) = \mathbb{P}\left( {{\rm{EE}} \ge \tau } \right),
\end{equation}
where $\tau$ is the target EE threshold. To derive the EEC, the user association probability, the SINR and rate coverages are required, which will be calculated in the following thoroughly.
\subsection{User Association Probabilities}\label{sec:4a}
It is known that the users are randomly distributed spatially in the network, and therefore, any given user can be divided into different user sets defined in~\eqref{eq3} with certain probability.

\vspace{2mm}

\begin{lem}\label{Lem:1}
\textit{Based on the definition of the three disjoint user sets in~\eqref{eq3}, the user association probability, defined as ${A_l} = \mathbb{P}\left( {u \in {u_l}} \right),l \in \left\{ {1,D,\bar D} \right\}$, can be derived by}
\begin{equation}\label{eq5}
{A_1} \hspace{-1mm} = \hspace{-1mm} 2\pi {\lambda _1} \hspace{-1mm} \int_0^\infty \hspace{-1mm} \hspace{-1mm} {r\exp \left( { - \pi {\lambda _1}{r^2} \hspace{-1mm} - \hspace{-1mm} \pi {\lambda _2}{{\left( {\frac{{{P_2}{B_1}}}{{{P_1}}}} \right)}^{ \hspace{-1mm} \frac{2}{\alpha _2}}}{r^{\frac{2 \alpha _1}{\alpha _2}}}} \right)} dr,
\end{equation}
\begin{equation}\label{eq6}
{A_{\bar D}} \hspace{-1mm} = \hspace{-1mm} 2\pi {\lambda _2} \hspace{-1mm} \int_0^\infty \hspace{-1mm} \hspace{-1mm} {r\exp \left( { - \pi {\lambda _2}{r^2} \hspace{-1mm} - \hspace{-1mm} \pi {\lambda _1}{{\left( {\frac{{{P_1}}}{{{P_2}{B_2}}}} \right)}^{ \hspace{-1mm} \frac{2}{\alpha _1}}} \hspace{-1mm} {r^{\frac{2 \alpha _2}{\alpha _1}}}} \right)} dr,
\end{equation}
\begin{equation}\label{eq7}
\begin{aligned}
 {A_D} \hspace{-1mm} = & 2\pi {\lambda _2} \hspace{-1mm} \int_0^\infty \hspace{-1mm} \hspace{-1mm} {r\left\{ {\exp \left( { - \pi {\lambda _2}{r^2} \hspace{-1mm} - \hspace{-1mm} \pi {\lambda _1}{{\left( {\frac{{{P_1}}}{{{P_2}{B_1}}}} \right)}^{ \hspace{-1mm} \frac{2}{\alpha _1}}} \hspace{-1mm} {r^{\frac{2 \alpha _2}{\alpha _1}}}} \right)} \right.} \\
 & \left. { - \exp \left( { - \pi {\lambda _2}{r^2} - \pi {\lambda _1}{{\left( {\frac{{{P_1}}}{{{P_2}{B_2}}}} \right)}^{\frac{2}{\alpha _1}}}{r^{\frac{2 \alpha _2}{\alpha _1}}}} \right)} \right\}dr.
\end{aligned}
\end{equation}
\end{lem}

\vspace{2mm}

\begin{IEEEproof}
See Appendix \ref{Apx1}.
\end{IEEEproof}
\subsection{SINR Coverage}\label{sec:4b}
The SINR coverage probability of a typical user $i$ in set ${{u_l}}$ for a target SINR threshold $T$ is defined as ${\rm{S}}\left( T \right){\rm{ = }}\mathbb{P}\left( {{\rm{SIN}}{{\rm{R}}_i} \ge T{\rm{|}}i \in {u_l}} \right)$~\cite{ref21}. For a typical user, the SINR includes the information of the distance from its serving BS, denoted by $G_l,l \in \left\{ {1,D,\bar D} \right\}$. Since the location of the BS obeys the Poisson distribution, $G_l$ is a random variable, based on which we need to attain the probability density function~(PDF) of $G_l$ first before we obtain the SINR coverage.

\vspace{2mm}

\begin{lem}\label{Lem:6}
\textit{The PDF of the distance between a typical user and its serving BS, i.e., $G_l$, can be formulated as}
\begin{equation}\label{eq6:1}
{f_{{G_1}}} \hspace{-1mm} \left( x \right) \hspace{-0.5mm} = \hspace{-0.5mm} \frac{{2\pi {\lambda _1}}}{{{A_1}}}x\exp \hspace{-0.5mm} \left( \hspace{-0.5mm} { - \pi {\lambda _1}{x^2} \hspace{-0.5mm} - \hspace{-0.5mm} \pi {\lambda _2}{{\left( \hspace{-0.5mm} {\frac{{{P_2}{B_1}}}{{{P_1}}}} \hspace{-0.5mm} \right)}^{ \hspace{-1mm} \frac{2}{{{\alpha _2}}}}} \hspace{-1mm} {x^{\frac{{2{\alpha _1}}}{{{\alpha _2}}}}}} \hspace{-0.5mm} \right),
\end{equation}
\begin{equation}\label{eq6:2}
{f_{{G_{\bar D}}}} \hspace{-1mm} \left( x \right) \hspace{-0.5mm} = \hspace{-0.5mm} \frac{{2\pi {\lambda _2}}}{{{A_{\bar D}}}}x\exp \hspace{-0.5mm} \left( \hspace{-1mm} { - \pi {\lambda _1}{{\left( \hspace{-0.5mm} {\frac{{{P_1}}}{{{P_2}{B_2}}}} \hspace{-0.5mm} \right)}^{ \hspace{-1mm} \frac{2}{{{\alpha _1}}}}}\hspace{-1mm}{x^{\frac{{2{\alpha _2}}}{{{\alpha _1}}}}} \hspace{-1mm} - \hspace{-1mm} \pi {\lambda _2}{x^2}} \hspace{-0.5mm} \right),
\end{equation}
\begin{equation}\label{eq6:3}
\begin{aligned}
 {f_{{G_D}}} \hspace{-1mm} \left( x \right) \hspace{-0.5mm} = \hspace{-0.5mm} & \frac{{2\pi {\lambda _2}}}{{{A_D}}}x \hspace{-0.5mm} \left\{ \hspace{-0.5mm} {\exp \hspace{-0.5mm} \left( \hspace{-0.5mm} { - \pi {\lambda _1}{{ \hspace{-0.5mm} \left( \hspace{-0.5mm} {\frac{{{P_1}}}{{{P_2}{B_1}}}} \hspace{-0.5mm} \right)}^{\hspace{-1mm} \frac{2}{{{\alpha _1}}}}} \hspace{-1mm} {x^{\frac{{2{\alpha _2}}}{{{\alpha _1}}}}} \hspace{-1mm} - \hspace{-1mm} \pi {\lambda _2}{x^2}} \hspace{-0.5mm} \right)} \right. \\
 & \left. { - \exp \hspace{-0.5mm} \left( \hspace{-0.5mm} { - \pi {\lambda _1}{{\left( {\frac{{{P_1}}}{{{P_2}{B_2}}}} \right)}^{\hspace{-1mm} \frac{2}{{{\alpha _1}}}}} \hspace{-1mm} {x^{\frac{{2{\alpha _2}}}{{{\alpha _1}}}}} \hspace{-0.5mm} - \hspace{-0.5mm} \pi {\lambda _2}{x^2}} \hspace{-0.5mm} \right)} \hspace{-0.5mm} \right\}.
\end{aligned}
\end{equation}
\end{lem}

\vspace{2mm}

\begin{IEEEproof}
See Appendix \ref{Apx2}.
\end{IEEEproof}

\vspace{2mm}

Based on Lemma~\ref{Lem:6}, we can further calculate the SINR coverage for a typical user as the following theorem.

\vspace{2mm}

\begin{theorem}\label{Lem:2}
\textit{For a typical user ${i \in {u_l}}$, the SINR coverage ${{{S}}_l}$ can be formulated as~\eqref{eq8}--\eqref{eq11} for $l \in \left\{ {B,\bar B,D,\bar D} \right\}$.}

\newcounter{TempEqCnt}
\setcounter{TempEqCnt}{\value{equation}}
\setcounter{equation}{11}
\begin{figure*}[htbp]
%\hrulefill
\begin{equation}\label{eq8}
{S_B}\left( T \right) = \frac{{2\pi {\lambda _1}}}{{{A_1}}}\int_{x = 0}^\infty  {x\exp } \left\{ { - \frac{{{x^{{\alpha _1}}}T{\sigma ^2}}}{{\beta {P_1}}} - \pi {\lambda _1}{x^2}\left[ {1 + Q \left( {T,{\alpha _1}} \right)} \right]} \right. - \left. {\pi {\lambda _2}{x^{\frac{{2{\alpha _1}}}{{{\alpha _2}}}}}{{\left( {\frac{{{P_2}{B_1}}}{{{P_1}}}} \right)}^{\frac{2}{{{\alpha _2}}}}}\left[ {1 + Q \left( {\frac{T}{{\beta {B_1}}},{\alpha _2}} \right)} \right]} \right\}dx,
\end{equation}

%\vspace{-1em}

\begin{equation}\label{eq9}
{S_{\bar B}}\left( T \right) = \frac{{2\pi {\lambda _1}}}{{{A_1}}}\int_{x = 0}^\infty  {x\exp } \left\{ { - \frac{{{x^{{\alpha _1}}}T{\sigma ^2}}}{{{P_1}}} - \pi {\lambda _1}{x^2}\left[ {1 + Q \left( {T,{\alpha _1}} \right)} \right]} \right.{\kern 1pt} {\kern 1pt}  - \left. {\pi {\lambda _2}{x^{\frac{{2{\alpha _1}}}{{{\alpha _2}}}}}{{\left( {\frac{{{P_2}{B_1}}}{{{P_1}}}} \right)}^{\frac{2}{{{\alpha _2}}}}}\left[ {1 + Q \left( {\frac{T}{{{B_1}}},{\alpha _2}} \right)} \right]} \right\}dx,
\end{equation}

%\vspace{-1em}
\begin{equation}\label{eq10}
\begin{aligned}
 {S_D}\left( T \right) & = \frac{{2\pi {\lambda _2}}}{{{A_D}}}\int_{x = 0}^\infty  {x\left\{ {\exp \left( { - \frac{{{x^{{\alpha _2}}}T{\sigma ^2}}}{{{P_2}}} - \pi {\lambda _2}{x^2}\left[ {1 + Q \left( {T,{\alpha _2}} \right)} \right] - \pi {\lambda _1}{x^{\frac{{2{\alpha _2}}}{{{\alpha _1}}}}}{{\left( {\frac{{{P_1}}}{{{P_2}{B_1}}}} \right)}^{\frac{2}{{{\alpha _1}}}}}\left[ {1 + Q \left( {\beta T{B_1},{\alpha _1}} \right)} \right]} \right)} \right.} \\
 & \left. { \hspace{-1mm} - \hspace{-0.5mm} \exp \hspace{-1mm} \left( \hspace{-0.5mm} { - \frac{{{x^{{\alpha _2}}}T{\sigma ^2}}}{{{P_2}}} \hspace{-0.5mm} - \hspace{-0.5mm} \pi {\lambda _2}{x^2}\left[ {1 \hspace{-0.5mm} + \hspace{-0.5mm} Q \left( {T,{\alpha _2}} \right)} \right] \hspace{-0.5mm} - \hspace{-0.5mm} \pi {\lambda _1}{{\left( {\frac{{{P_1}}}{{{P_2}{B_2}}}} \right)}^{\frac{2}{{{\alpha _1}}}}}{x^{\frac{{2{\alpha _2}}}{{{\alpha _1}}}}}\left[ {1 \hspace{-0.5mm} + \hspace{-0.5mm} {{\left( {\beta T{B_2}} \right)}^{\frac{2}{{{\alpha _1}}}}} \hspace{-1mm} \int_{{{\left( {\beta T{B_1}} \right)}^{\frac{{{\rm{ - }}2}}{{{\alpha _1}}}}}}^\infty \hspace{-0.5mm} {\frac{1}{{1 \hspace{-0.5mm} + \hspace{-0.5mm} {v^{\frac{{{\alpha _1}}}{2}}}}}} dv} \right]} \hspace{-0.5mm} \right)} \right\}dx,
\end{aligned}
\end{equation}

%\vspace{-1em}

\begin{equation}\label{eq11}
{S_{\bar D}}\left( T \right) = \frac{{2\pi {\lambda _2}}}{{{A_{\bar D}}}}\int_{x = 0}^\infty  {x\exp } \left\{ { - \frac{{{x^{{\alpha _2}}}T{\sigma ^2}}}{{{P_2}}} - \pi {\lambda _2}{x^2}\left[ {1 + Q \left( {T,{\alpha _2}} \right)} \right]} \right. - \left. {\pi {\lambda _1}{x^{\frac{{2{\alpha _2}}}{{{\alpha _1}}}}}{{\left( {\frac{{{P_1}}}{{{P_2}{B_2}}}} \right)}^{\frac{2}{{{\alpha _1}}}}}\left[ {1 + Q \left( {T{B_2},{\alpha _1}} \right)} \right]} \right\}dx,
\end{equation}
where $Q \left( {T,\alpha } \right) = {T^{\frac{2}{\alpha }}}\int_{{T^{\frac{{ - 2}}{\alpha }}}}^\infty  {\frac{1}{{1 + {x^{\frac{\alpha }{2}}}}}} dx$.

\hrulefill
\end{figure*}

\setcounter{equation}{15}

\end{theorem}

\vspace{2mm}

\begin{IEEEproof}
See Appendix \ref{Apx3}.
\end{IEEEproof}

\vspace{2mm}

It is noted that the overall average SINR coverage is used as the metric to characterize the network coverage performance. Based on Lemma~\ref{Lem:1} and Theorem~\ref{Lem:2}, the overall average SINR coverage can be obtained, as given by
\begin{equation}\label{eq12}
S\left( T \right) = {A_1}{S_1}\left( T \right) + {A_D}{S_D}\left( T \right) + {A_{\bar D}}{S_{\bar D}}\left( T \right),
\end{equation}
where ${S_1}\left( T \right)$ is the conditional SINR coverage of a typical user in macro cells. As shown in Fig.~\ref{fig.1}, a macro user is scheduled in two types of resources, i.e., low-power and full-power transmit resources, with the probabilities of $\eta$ and $1 - \eta$, respectively, based on which we have
\begin{equation}\label{eq13}
{S_1}\left( T \right) = \eta {S_B}\left( T \right) + \left( {1 - \eta } \right){S_{\bar B}}\left( T \right).
\end{equation}

From Theorem~\ref{Lem:2}, it is observed that ${S_1}$ is only dependent on $B_1$, ${S_{\bar D}}$ is only dependent on $B_2$, and ${S_D}$ is dependent on both $B_1$ and $B_2$. This is because that the user set ${u_1}$ is determined by $B_1$, the user set ${u_{\bar D}}$ is determined by $B_2$, and the user set ${u_D}$ is determined by $B_1$ and $B_2$ jointly, as shown in~Fig.~\ref{fig.systemmodel}. Since $\eta$ represents the probability that a macro user is scheduled in the low-power transit resources, the SINR coverage~$S_1$ is dependent on~$\eta$, revealed in~\eqref{eq13}, however, $\eta$ has no effect on the individual SINR coverage $S_l$. Further, the transmit power reduction only affects the users in ${u_B}$ and ${u_D}$, so $\beta$ has impact on ${S_B}$ and ${S_D}$ rather than ${S_{\bar B}}$ and ${S_{\bar D}}$. Therefore, in summary, the overall SINR coverage~$S$ is dependent on the association biases $B_1$ and $B_2$, the resource partitioning fraction~$\eta$, and the transmit power fraction~$\beta$, which is consistent with our theoretical analysis.
\subsection{Data Rate Coverage}\label{sec:4c}
Like the SINR coverage, the data rate coverage of a typical user $i$ in user set ${{u_l}}$ for a target rate threshold $\rho$ is defined as $R\left( \rho  \right){\rm{ = }}\mathbb{P}\left( {{\rm{Rate}}_i \ge \rho {\rm{|}}i \in {u_l}} \right)$. In the data rate coverage derivation, the following mapping relationships are used:
\begin{equation}\nonumber
k\left( l \right) = \left\{ {\begin{array}{ll}
   \hspace{-2mm} 1, & {{\rm if}\ l \in \left\{ {B,\bar B,D} \right\}} \\
   \hspace{-2mm} 2, & {{\rm if}\ l \in \left\{ {\bar D} \right\}} \\
\end{array}} \right. \hspace{-3mm} ,\ q\left( l \right) = \left\{ {\begin{array}{ll}
   \hspace{-2mm} 1, & {{\rm if}\ l \in \left\{ {B,\bar B} \right\}} \\
   \hspace{-2mm} D, & {\rm if}\ l \in \left\{ D \right\} \\
   \hspace{-2mm} {\bar D}, & {\rm if}\ l \in \left\{ {\bar D} \right\} \\
\end{array}} \right. \hspace{-3mm} .
\end{equation}

Furthermore, we will introduce the probability mass function of the users associated with the tagged BS of a PPP~\cite{ref5,ref14}, denoted by ${p_l}\left( n \right)$, as expressed by
\begin{equation}\label{eq15}
\begin{aligned}
 {p_l}\left( n \right) = & \mathbb{P} \left( {{N_{q\left( l \right)}} = n} \right) \\
 = & \frac{{{{3.5}^{3.5}}}}{{n!}}\frac{{\Gamma \left( {n + 3.5} \right)}}{{\Gamma \left( {3.5} \right)}}{\left( {\frac{{{\lambda _u}{A_{q\left( l \right)}}}}{{{\lambda _{M\left( l \right)}}}}} \right)^{n - 1}} \times \\
 & {\left( {3.5 + \frac{{{\lambda _u}{A_{q\left( l \right)}}}}{{{\lambda _{M\left( l \right)}}}}} \right)^{ - \left( {n + 3.5} \right)}},\ \left( {n \ge 1} \right),
\end{aligned}
\end{equation}
where ${{N_{q\left( l \right)}}}$ characterizes the load at the tagged BS and $\Gamma \left( M \right) = \int_0^\infty  {{t^{M - 1}}} \exp \left( { - t} \right)dt$ is the gamma function.

With the help of~\eqref{eq15}, we can calculate the data rate coverage for a typical user as the following theorem.

\vspace{2mm}

\begin{theorem}\label{Lem:3}
\textit{For a typical user ${i \in {u_l}}$, the data rate coverage ${{{R}}_l}$ can be formulated for $l \in \left\{ {B,\bar B,D,\bar D} \right\}$, as given by}
\begin{equation}\label{eq14}
\begin{aligned}
 {R_l} \left({\rho}\right)  = & \sum\limits_{n \ge 1} \frac{{{{3.5}^{3.5}}}}{{n!}}\frac{{\Gamma \left( {n + 3.5} \right)}}{{\Gamma \left( {3.5} \right)}}{{\left( {\frac{{{\lambda _u}{A_{q\left( l \right)}}}}{{{\lambda _{M\left( l \right)}}}}} \right)}^{n - 1}} \times \\
 & {{\left( {3.5 + \frac{{{\lambda _u}{A_{q\left( l \right)}}}}{{{\lambda _{M\left( l \right)}}}}} \right)}^{ - \left( {n + 3.5} \right)}} {S_l}\left( {{2^{\frac{{{\rho _{k\left( l \right)}}n}}{W}}} - 1} \right).
\end{aligned}
\end{equation}
\end{theorem}

\vspace{2mm}

\begin{IEEEproof}
According to Shannon theorem, the data rate coverage ${{{R}}_l}$ for a typical user ${i \in {u_l}}$ can be formulated as
\begin{equation}\label{eq16}
\begin{aligned}
 {R_l} \left({\rho}\right) & = \mathbb{P}\left( {{\rm Rate}_l \ge {\rho _{k\left( l \right)}}} \right) \\
 & = \mathbb{P} \left( {\frac{W}{{{N_{q\left( l \right)}}}}{{\log }_2}\left( {1 + {\rm{SINR}}} \right) \ge {\rho _{k\left( l \right)}}} \right) \\
 & = \mathbb{P}\left( {{\rm{SINR}} \ge {2^{\frac{{{\rho _{k\left( l \right)}}{N_{q\left( l \right)}}}}{W}}} - 1} \right) \\
 & = \mathbb{E}{_{{N_{q\left( l \right)}}}}\left[ {{S_l}\left( {{2^{\frac{{{\rho _{k\left( l \right)}}{N_{q\left( l \right)}}}}{W}}} - 1} \right)} \right] \\
 & \mathop {\rm{ = }}\limits^{\left( a \right)} \sum\limits_{n \ge 1} {{p_l}} \left( n \right){S_l}\left( {{2^{\frac{{{\rho _{k\left( l \right)}}n}}{W}}} - 1} \right),
\end{aligned}
\end{equation}
where $\left( a \right)$ ignores the dependence between the load and the SINR coverage for tractability of the analysis~\cite{ref14}.

Combining \eqref{eq15} and \eqref{eq16}, Eq.~\eqref{eq14} can be obtained consequently.
\end{IEEEproof}

\vspace{2mm}

The data rate coverage~\eqref{eq14} can be further simplified (sacrificing accuracy) if the load at each BS is assumed to be its mean. Therefore, instead of considering the cell size distribution and the user distribution in each cell, we adopt the mean load approximation proposed in~\cite{ref14}, where the mean cell load ${{N_{q\left( l \right)}}}$ is given by
\begin{equation}\label{eq17}
{\bar N_{q\left( l \right)}} = 1 + \frac{{1.28{\lambda _u}{A_{q\left( l \right)}}}}{{{\lambda _{M\left( l \right)}}}},
\end{equation}
based on which the following proposition is proposed.

\vspace{2mm}

\begin{proposition}\label{Lem:4}
\textit{Based on the approximated cell load, the data rate coverage ${{{R}}_l}$ for a typical user ${i \in {u_l}}$, $l \in \left\{ {B,\bar B,D,\bar D} \right\}$ can be reformulated as}

\begin{equation}\label{eq18}
 {R_l} \left({\rho}\right) = {S_l}\left( {{2^{{{{\rho _{k\left( l \right)}}{(1 + \frac{{1.28{\lambda _u}{A_{q\left( l \right)}}}}{{{\lambda _{M\left( l \right)}}}})}}/W}}} - 1} \right).
\end{equation}
\end{proposition}

\vspace{2mm}

\begin{IEEEproof}
According to~\eqref{eq16}, ${R_l} \left({\rho}\right)$ can be expressed as
\begin{equation}\label{eq23}
\begin{aligned}
 {R_l} \left({\rho}\right) & = \mathbb{E}{_{{N_{q\left( l \right)}}}}\left[ {{S_l}\left( {{2^{\frac{{{\rho _{k\left( l \right)}}{N_{q\left( l \right)}}}}{W}}} - 1} \right)} \right] \\
 & = {S_l}\left( {{2^{\frac{{{\rho _{k\left( l \right)}}\mathbb{E}\left[ {{N_{q\left( l \right)}}} \right]}}{W}}} - 1} \right) \\
 & = {S_l}\left( {{2^{\frac{{{\rho _{k\left( l \right)}}{{\bar N}_{q\left( l \right)}}}}{W}}} - 1} \right).
\end{aligned}
\end{equation}
By substituting~\eqref{eq17} into~\eqref{eq23}, Eq.~\eqref{eq18} can be attained.
\end{IEEEproof}

\vspace{2mm}

Being similar to the calculation of overall average SINR coverage, the overall average data rate coverage is
\begin{equation}\label{eq19}
R\left( \rho  \right) = {A_1}{R_1}\left( \rho  \right) + {A_D}{R_D}\left( \rho  \right) + {A_{\bar D}}{R_{\bar D}}\left( \rho  \right),
\end{equation}
where ${R_1}\left( \rho  \right) = \eta {R_B}\left( \rho  \right) + \left( {1 - \eta } \right){R_{\bar B}}\left( \rho  \right)$.
\subsection{Energy Efficiency Coverage}\label{sec:4d}
Utilizing the data rate coverage, the EE coverage can be obtained with some mathematical transformations, as presented by the following theorem.

\vspace{2mm}

\begin{theorem}\label{Lem:5}
\textit{For a typical user ${i \in {u_l}}$, the EE coverage ${{\rm{C}}_l}$ can be formulated for $l \in \left\{ {B,\bar B,D,\bar D} \right\}$, as given by}
\begin{equation}\label{eq20}
{C_l}\left( \tau  \right) =\mathbb{P} \left( {E{E_l} \ge \tau } \right) = {R_l}\left( {\tau {P_{M\left( l \right),{\rm total}}}} \right),
\end{equation}
\textit{where ${R_l}$ can be obtained from Theorem~\ref{Lem:3} or Proposition~\ref{Lem:4}.}
\end{theorem}

\vspace{2mm}

\begin{IEEEproof}
According to the definition of EEC, the EE coverage ${{\rm{C}}_l}$ for $l \in \left\{ {B,\bar B,D,\bar D} \right\}$ can be formulated as
\begin{equation}\label{eq21}
\begin{aligned}
 {C_l}\left( \tau  \right) &= \mathbb{P}\left( {E{E_l} \ge \tau } \right) \\
 &= \mathbb{P} \left( {\frac{{{R_l}}}{{{P_{M\left( l \right),{\rm total}}}}} \ge \tau } \right) \\
 &= \mathbb{P} \left( {{R_l} \ge \tau {P_{M\left( l \right),{\rm total}}}} \right) \\
 &= {R_l}\left( {\tau {P_{M\left( l \right),{\rm total}}}} \right),
\end{aligned}
\end{equation}
which then can be calculated by the data rate coverage presented in Theorem~\ref{Lem:3} or Proposition~\ref{Lem:4}.
\end{IEEEproof}

\vspace{2mm}

Then, the overall average EE coverage can be calculated as
\begin{equation}\label{eq22}
C\left( \tau  \right) = {A_1}{C_1}\left( \tau  \right) + {A_D}{C_D}\left( \tau  \right) + {A_{\bar D}}{C_{\bar D}}\left( \tau  \right),
\end{equation}
where ${C_1}\left( \tau  \right) = \eta {C_B}\left( \tau  \right) + \left( {1 - \eta } \right){C_{\bar B}}\left( \tau  \right)$. It is seen that the close-form expression of EE coverage has not been derived, but only the simple numerical integration is involved which is very easy to calculate to any desired accuracy using numerical integration methods with computer in practice.
\section{Numerical Results and Discussion}\label{sec:5}
In this section, simulations are carried out to show the benefits of the extra CRE and the effects of different parameters on SINR, data rate, and EE coverages respectively, where the simulation parameters are listed in Table~\ref{tab:1}. Moreover, the traditional CRE is also conducted for comparison, which is simulated by setting ${B_1} = {B_2}$ in each situation.

\begin{table}[b]
\vspace{-3mm}
\centering
\caption{Simulation Parameters}
\vspace{-2mm}
\begin{tabular}{l|l}
\hline
\textbf{Parameters} & \textbf{Values} \\
\hline
System bandwidth ($W$) & 10\,MHz \\
Noise power spectral density ($N_0$) & --\,174\,dBm/Hz \\
The density of BSs (${\lambda _1}$, ${\lambda _2}$) & 1, 10\,BS/$\rm km^2$ \\
The density of users (${\lambda _u}$) & 100\,user/$\rm km^2$ \\
The transmit power ($P_1$, $P_2$) & 10, 0.1\,W \\
Path loss factor ($\alpha _1$, $\alpha _2$) & 3.5, 4.0 \\
The target data rate ($\rho _1$, $\rho _2$) & 300, 1200\,kbps \\
Power parameters of macro BS ($a_1$, $b_1$) & 22.6, 412.4 \\
Power parameters of small cell BS ($a_2$, $b_2$) & 5.5, 32 \\
\hline
\end{tabular}
\label{tab:1}
%\vspace{-3mm}
\end{table}

\begin{figure}[!t]%[htbp]
%\vspace{-0.3cm}
\centering
\includegraphics[width=3.5in]{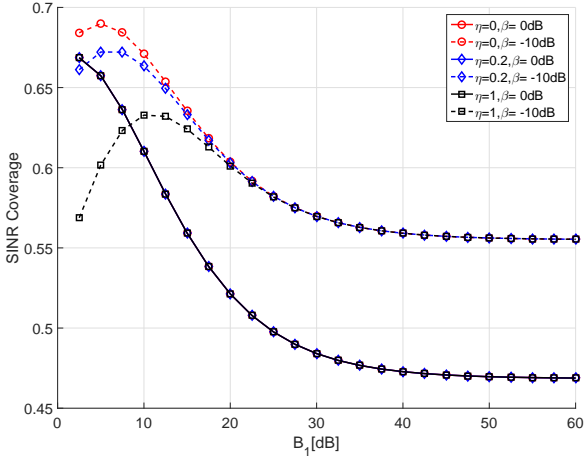}
\vspace{-8mm}
\caption{Effect of association bias $B_1$ on the SINR coverage (${B_2} = 2.5$ dB).}
\label{fig.2}
\vspace{-3mm}
\end{figure}
Fig.~\ref{fig.2} shows the effect of the association bias $B_1$ on the SINR coverage, when ${B_2} = 2.5$ dB. The traditional CRE is achieved at the start point of the curves, i.e., ${B_1} = 2.5$ dB. It is obvious that the extra CRE with transmit power reduction can definitely improve the SINR coverage of the network. With transmit power reduction, it is observed that the SINR coverage first increases and then decreases as $B_1$ increases; in other words, the optimal association bias $B_1$ which is not equal to $B_2$ can be found when $\beta < 0$ dB. This is because that $B_1$ determines how many original macro users with data rate $\rho_1$ will be offloaded to small cells. As $B_1$ increases at first, more original macro users with relatively short distance will be served by small cells, causing little received power degradation, but the interference of the extra offloaded users can be managed and controlled by transmit power reduction of macro BSs, which overall contributes to the average SINR coverage. However, when $B_1$ exceeds a certain threshold, due to the continuing grown of set ${u_{\bar D}}$, which contains more macro users having long distance away from small cells, the average received power of users in ${u_{\bar D}}$ becomes so worse that cannot be compensated by transmit power reduction, resulting in the decrease of the overall average SINR coverage.

Unlike the reduced power transmission case, the SINR coverage is a non-increasing function of $B_1$ with full power transmission (i.e., $\beta = 0$ dB), as shown in Fig.~\ref{fig.2}. It is clear that without transmit power reduction, the users in the extra cell expanded area will suffer poor received power and great interference, even without any compensation. Furthermore, without transmit power reduction, the SINR coverage is independent of the partitioning fraction $\eta$, because the users in the macro cells suffer the same SINR on the two types of resources. It is also seen that the SINR coverage degrades since the partitioning fraction $\eta$ increases with low power transmission. It is because that the partitioning fraction $\eta$ only influences the conditional SINR coverage of a typical user in macro cell, and $\eta$ always decreases the SINR coverage of macro users as ${\rm{SIN}}{{\rm{R}}_B}$ is always smaller than ${\rm{SIN}}{{\rm{R}}_{\bar B}}$.

\begin{figure}[!t]%[htbp]
%\vspace{-0.3cm}
\centering
\includegraphics[width=3.5in]{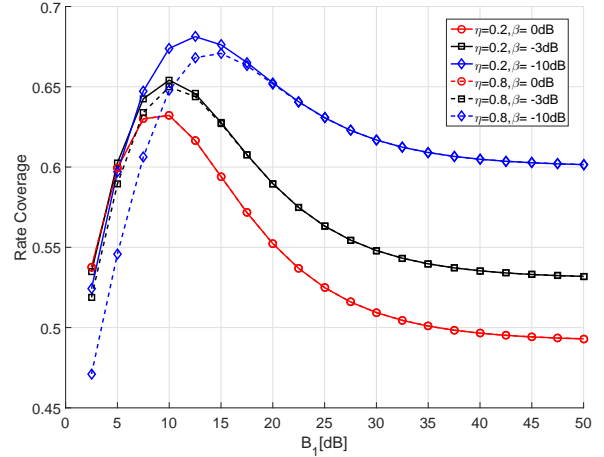}
\vspace{-8mm}
\caption{Effect of association bias $B_1$ on the rate coverage (${B_2} = 2.5$ dB).}
\label{fig.3}
\vspace{-3mm}
\end{figure}
Fig.~\ref{fig.3} illustrates the effect of the association bias $B_1$ on the data rate coverage, when $B_2 = 2.5$ dB. Compared with the traditional CRE, i.e., the start point of the curves, it can be seen that the extra CRE can improve the data rate coverage of the networks, and the optimal $B_1$ of extra CRE can be found for all cases. Moreover, the data rate coverage first increase and then decrease until the saturation with the association bias $B_1$ increasing for all curves. It is because that as mentioned before $B_1$ determines how many original macro users will be offloaded to small cells, and as $B_1$ increases initially, more original macro users will be served by small cells to balance the traffic load and take full use of small cells, which improves the data rate coverage. However, when the extra CRE becomes too larger, much more users will connect to small cells, which makes the small cells congested. Finally, all users will be served by small cells, so the data rate coverage tends unchanged. Similarly, without transmit power reduction, the data rate coverage is independent of the partitioning fraction $\eta$, where the reason is that $\eta$ only influences the conditional rate coverage of a typical user in macro cells.

\begin{figure}[!t]%[htbp]
%\vspace{-0.3cm}
\centering
\includegraphics[width=3.5in]{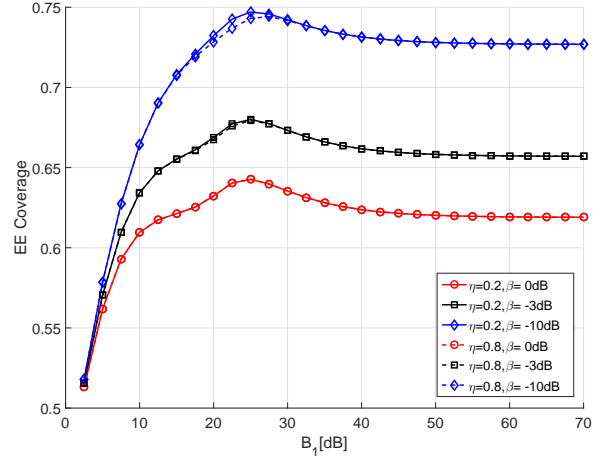}
\vspace{-8mm}
\caption{Effect of association bias $B_1$ on the EE coverage (${B_2} = 2.5$ dB).}
\label{fig.9}
\vspace{-3mm}
\end{figure}
Fig.~\ref{fig.9} plots the effect of the association bias $B_1$ on the EE coverage, when $B_2 = 2.5$ dB. Note that the start point of the curves, i.e., ${B_1} = {B_2} = 2.5$ dB, is the traditional CRE, which can be interpreted as the benchmark. It is clear that the extra CRE can promote the EE coverage of the network, and there is an optimal $B_1$ of extra CRE for all cases. It is seen that with transmit power reduction, the EE coverage always outperforms the case without transmit power reduction (i.e. $\beta = 0$ dB) at any given $B_1$, which indicates there is an improvement of EE as a whole. Therefore, it is reasonable to say that low power transmission provides superiority in EE performance of the HCN. Moreover, it can be observed that without transmit power reduction, the EE coverage is also independent of the partitioning fraction $\eta$, where the reason has been presented in analysis of SINR and data rate coverages.

\begin{figure}[!t]%[htbp]
%\vspace{0.12cm}
\centering
\includegraphics[width=3.5in]{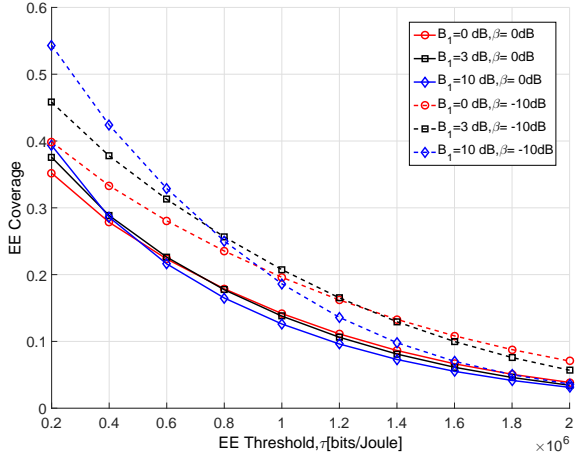}
\vspace{-8mm}
\caption{EE coverage for different $B_1$ and $\beta$ ($B_2 = 0$ dB, $\eta = 0.2$).}
\label{fig.5}
\vspace{-3mm}
\end{figure}
In Fig.~\ref{fig.5}, we first set the association bias $B_1$ to different values and plot the EE coverage curves together for comparison, where $B_2 = 0$ dB and $\eta = 0.2$. It is clear that the EE coverage improvement can be found between these different schemes when the EE threshold is smaller than a certain value, and the bigger $B_1$ is, the better the EE coverage will be. However, as the EE threshold increases, it can significantly reverse the relationship between the EE coverage and the association bias $B_1$, i.e., a transition point can be found when compare the curves with different $B_1$. Furthermore, with transmit power reduction, the transition point of the EE threshold is bigger than that with full transmit power, which means that from the perspective of EE performance, compared with the case without transmit power reduction, more macro users that are at the edge of macro cells can be offloaded to the small cells with transmit power reduction.

\begin{figure}[!t]%[htbp]
%\vspace{0.32cm}
\centering
\includegraphics[width=3.5in]{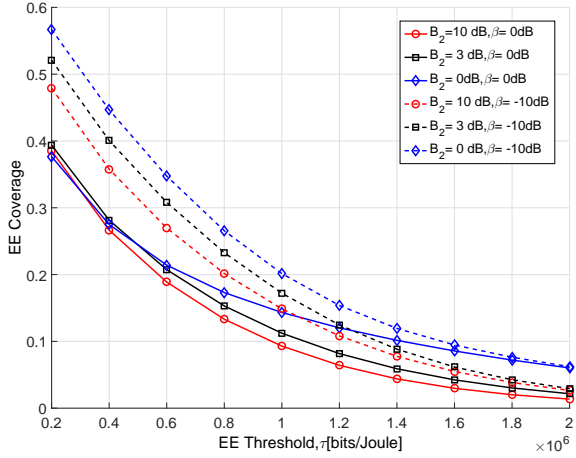}
\vspace{-8mm}
\caption{EE coverage for different $B_2$ and $\beta$ ($B_1 = 10$ dB, $\eta = 0.2$).}
\label{fig.6}
\vspace{-3mm}
\end{figure}
In Fig.~\ref{fig.6}, we set the association bias $B_2$ to different values and plot the EE coverage curves together for comparison, where $B_1 = 10$ dB and $\eta = 0.2$. From this figure, it can be seen that there is also an EE coverage improvement between these different schemes, and the smaller $B_2$ is, the better the EE coverage will be. It is easy to draw this conclusion since bigger $B_2$ means more users with worse channel condition are transferred to the set ${u_{\bar D}}$ and less extra macro users are offloaded to small cells with reduced interference, which all result in the degradation of the overall SINR and data rate coverages, and therefore, influence the EE coverage consequently. For the special case where ${B_2} = {B_1}$ which means ${u_D} = \emptyset $, the EE coverage will achieve the worst value. The reason is that there is no extra offloaded users and no performance gain brought by transmit power reduction in this case, and therefore leads to the worst EE coverage.

\begin{figure}[!t]%[htbp]
%\vspace{0.32cm}
\centering
\includegraphics[width=3.5in]{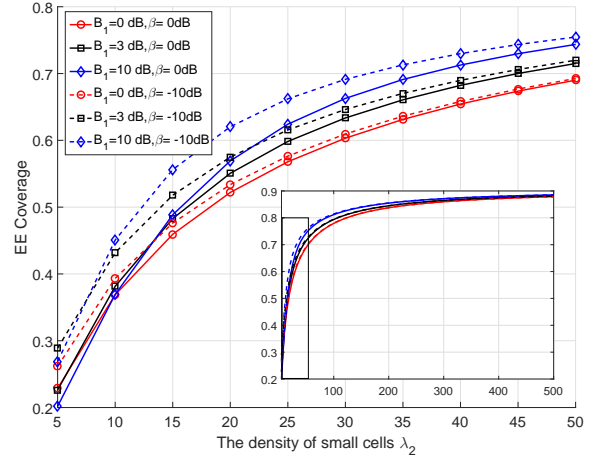}
\vspace{-8mm}
\caption{Effect of the density of small cells on EE coverage with different $B_1$ and $\beta$ ($B_2 = 0$ dB, $\eta = 0.2$).}
\label{fig.7}
\vspace{-3mm}
\end{figure}

Fig.~\ref{fig.7} shows the effect of the density of small cells on EE coverage with different $B_1$ and $\beta$, where $B_2 = 0$ dB and $\eta = 0.2$. It is observed that the EE coverage will become better but the improvement gradually becomes saturated, with the increase of the density of small cells. The reason is that with deploying more small cells, more users can be offloaded and the total traffic in the networks can be shared and balanced by more access points, which can definitely improve the system capacity as well as the EE performance. Moreover, with more small cells in the network, each user will have the much better channel condition due to the shorter distance away from its serving BS, which can significantly increase the SINR coverage. However, when the density exceeds a threshold and becomes too large, the total energy consumption will increase swiftly while the system capacity tends to increase slowly due to more interference introduced in the network, based on which the EE coverage will become almost stable. It also can seen in this figure that the extra CRE and transmit power reduction can benefit for the EE coverage of the network by comparing these curves.
\section{Conclusion and Future Work}\label{sec:6}
In this paper, we develop a general framework of extra CRE with joint transmit power reduction and resource partitioning in HCN, and analyze its system performance, including SINR, data rate, and EE coverages. Using stochastic geometry, we formulate the theoretical EE coverage as well as the SINR and data rate coverages. Simulations are carried out to verify our theoretical analysis, and the results reveal that the extra CRE and deploying more small cells (its benefit tends to be saturated) can both improve the EE of HCN. Several insights are obtained, which can provide valuable guidelines to practical design of future networks, especially to the traffic offloading in HCN. It is concluded that the parameters for extra CRE in HCN need to be optimized to acquire better system performance, which will be studied in our future works.

\appendices
\section{Proof of {Lemma}~\ref{Lem:1}}\label{Apx1}
\begin{IEEEproof}
Using the definition of the three disjoint user sets in~\eqref{eq3}, the user association probability ${A_1}$ can be derived as
\begin{equation}\label{Apx1:eq1}
\begin{aligned}
 {A_1} &= \mathbb{P} \left( {{P_1}D_1^{ - {\alpha _1}} \ge {P_2}{B_1}D_2^{ - {\alpha _2}}} \right) \\
 &= \mathbb{P} \left( {{D_2} \ge {{\left( {\frac{{{P_2}{B_1}}}{{{P_1}{D_1}^{ - {\alpha _1}}}}} \right)}^{\frac{1}{{{\alpha _2}}}}}} \right) \\
 &= \int_0^\infty \mathbb{P} \left( {{D_2} \ge {{\left( {\frac{{{P_2}{B_1}}}{{{P_1}{r^{ - {\alpha _1}}}}}} \right)}^{\frac{1}{{{\alpha _2}}}}}} \right){f_{{D_1}}}\left( r \right)dr \\
 &\mathop { = }\limits^{(a)} \int_0^\infty  {\exp \left( { - \pi {\lambda _2}{{\left( {\frac{{{P_2}{B_1}}}{{{P_1}{r^{ - {\alpha _1}}}}}} \right)}^{\frac{2}{{{\alpha _2}}}}}} \right)} {f_{{D_1}}}\left( r \right)dr,
\end{aligned}
\end{equation}
where (a) is because of the null probability of two-dimensional Poisson process. Similarly, ${A_{\bar D}}$ and ${A_D}$ can be presented by
\begin{equation}\label{Apx1:eq2}
 {A_{\bar D}} = \int_0^\infty  {\exp \left( { - \pi {\lambda _1}{{\left( {\frac{{{P_1}}}{{{P_2}{B_2}{r^{ - {\alpha _2}}}}}} \right)}^{\frac{2}{{{\alpha _1}}}}}} \right)} {f_{{D_2}}}\left( r \right)dr,
\end{equation}
\begin{equation}\label{Apx1:eq3}
\begin{aligned}
 {A_D} &= \int_0^\infty  {\exp \left( { - \pi {\lambda _1}{{\left( {\frac{{{P_1}}}{{{P_2}{B_1}{r^{ - {\alpha _2}}}}}} \right)}^{\frac{2}{{{\alpha _1}}}}}} \right)} {f_{{D_2}}}\left( r \right)dr \\
 &- \int_0^\infty  {\exp \left( { - \pi {\lambda _1}{{\left( {\frac{{{P_1}}}{{{P_2}{B_2}{r^{ - {\alpha _2}}}}}} \right)}^{\frac{2}{{{\alpha _1}}}}}} \right)} {f_{{D_2}}}\left( r \right)dr.
\end{aligned}
\end{equation}

Then we will calculate ${f_{{D_k}}}(k=1,2)$ which is involved in the expressions of ${A_1}$, ${A_{\bar D}}$ and ${A_D}$. Specifically, $\mathbb{P} \left( {{D_k} > r} \right)$ is derived by using the null probability of a two-dimensional Poisson process with the density of ${\lambda _k}$ in the area $A = \pi {r^2}$, which is expressed as $\exp \left( { - \pi {\lambda _k}{r^2}} \right)$. Since ${F_{{D_k}}}\left( r \right) = \mathbb{P} \left( {{D_k} < r} \right)$, ${f_{{D_k}}}\left( r \right)$ can be given by
\begin{equation}\label{Apx1:eq4}
\begin{aligned}
 {f_{{D_k}}}\left( r \right) &= \frac{{d\left( {1 - \mathbb{P} \left( {{D_k} > r} \right)} \right)}}{{dr}} \\
 &= \frac{{d\left( {1 - \exp \left( { - \pi {\lambda _k}{r^2}} \right)} \right)}}{{dr}} \\
 &= \exp \left( { - \pi {\lambda _k}{r^2}} \right)2\pi {\lambda _k}r.
\end{aligned}
\end{equation}

By substituting \eqref{Apx1:eq4} into \eqref{Apx1:eq1}--\eqref{Apx1:eq2}, Eqs. \eqref{eq5}--\eqref{eq7} are formulated, i.e., the lemma is completely proved.
\end{IEEEproof}

\section{Proof of {Lemma}~\ref{Lem:6}}\label{Apx2}
\begin{IEEEproof}
Given the association of the typical user with the $M\left( l \right)$-th tier, the event ${G_l} > x$ is equivalent to the event ${D_{M\left( l \right)}} > x$, whose probability can be formulated as
\begin{equation}\label{Apx2:eq1}
\begin{aligned}
\mathbb{P} \left( {{G_l} > x} \right) &= \mathbb{P} \left( {{D_{M\left( l \right)}} > x|u \in {u_l}} \right) \\
&= \frac{{\mathbb{P} \left( {{D_{M\left( l \right)}} > x,u \in {u_l}} \right)}}{{\mathbb{P} \left( {u \in {u_l}} \right)}} \\
&= \frac{{\mathbb{P} \left( {{D_{M\left( l \right)}} > x,u \in {u_l}} \right)}}{{A_l}},
\end{aligned}
\end{equation}
where ${G_B} = {G_{\bar B}} = {G_1}$ and ${A_B} = {A_{\bar B}} = {A_1}$. According to~\eqref{Apx2:eq1}, we take $\mathbb{P} \left( {{G_1} > x} \right)$ for example, which can be expressed as
\begin{equation}\label{Apx2:eq2}
\begin{aligned}
 \mathbb{P} \left( {{G_1} \hspace{-1mm} > \hspace{-1mm} x} \right) \hspace{-1mm} &= \hspace{-1mm} \frac{1}{{{A_1}}} \mathbb{P} \left( {{D_1} > x,u \in {u_1}} \right) \\
 \hspace{-1mm} &= \hspace{-1mm} \frac{1}{{{A_1}}} \mathbb{P} \left( {{D_1} > x,{P_1}D_1^{ - {\alpha _1}} \ge {P_2}{B_1}D_2^{ - {\alpha _2}}} \right) \\
 \hspace{-1mm} &= \hspace{-1mm} \frac{1}{{{A_1}}}\int_{r > x} \hspace{-2mm} {\mathbb{P} \left( {{D_2} \ge {{\left( {\frac{{{P_2}{B_1}}}{{{P_1}{r^{ - {\alpha _1}}}}}} \right)}^{\frac{1}{{{\alpha _2}}}}}} \right){f_{{D_1}}}\left( r \right)dr} \\
 \hspace{-1mm} &= \hspace{-1mm} \frac{{2\pi {\lambda _1}}}{{{A_1}}} \hspace{-2mm} \int_{r > x} \hspace{-4mm} {r\exp \left( \hspace{-1mm} { - \pi {\lambda _1}{r^2} \hspace{-1mm} - \hspace{-1mm} \pi {\lambda _2}{{\left( \hspace{-1mm} {\frac{{{P_2}{B_1}}}{{{P_1}}}} \hspace{-1mm} \right)}^{\frac{2}{{{\alpha _2}}}}} \hspace{-1mm} {r^{\frac{{2{\alpha _1}}}{{{\alpha _2}}}}}} \hspace{-1mm} \right)} dr.
\end{aligned}
\end{equation}
The cumulative distribution function~(CDF) of $G_1$ is ${F_{{G_1}}}\left( x \right) = 1 - \mathbb{P} \left( {{G_1} > x} \right)$, based on which the PDF of $G_1$ can be obtained, as given by \eqref{eq6:1}. Similarly, the PDF of $G_D$ and ${G_{\bar D}}$ can be formulated as \eqref{eq6:2} and \eqref{eq6:3}, respectively.
\end{IEEEproof}

\section{Proof of {Theorem}~\ref{Lem:2}}\label{Apx3}
\begin{IEEEproof}
For a typical user ${i \in {u_l}}$, $l \in \left\{ {B,\bar B,D,\bar D} \right\}$, the SINR coverage $S_l$ can be calculated as
\begin{equation}\label{Apx3:eq1}
{S_l}\left( T \right) = \int_{x = 0}^\infty  {\mathbb{P} \left[ {SIN{R_l}\left( x \right) \ge T} \right]} {f_{{G_l}}}\left( x \right)dx,
\end{equation}
where ${G_B} = {G_{\bar B}} = {G_1}$ and ${f_{{G_l}}}$ is obtained by {Lemma}~\ref{Lem:6}.

Using \eqref{eq4}, the CDF of the SINR of the user with the distance of $G_B$ can be formulated, as given by
\begin{equation}\label{Apx3:eq2}
\begin{aligned}
 &\mathbb{P} \left[ {SIN{R_B}\left( x \right) \ge T} \right] \\
 &= \mathbb{P} \left[ {\frac{{\beta {P_1}{h_x}{x^{ - {\alpha _1}}}}}{{\beta {I_{{r_1}}} + {I_{{r_2}}} + {\sigma ^2}}} \ge T} \right] \\
 &=\mathbb{P} \left[ {{h_x} \ge \frac{{{x^{{\alpha _1}}}T\left( {\beta {I_{{r_1}}} + {I_{{r_2}}} + {\sigma ^2}} \right)}}{{\beta {P_1}}}} \right] \\
 &\mathop  = \limits^{\left( a \right)} \mathbb{E}_{\left\{ {{I_{BB}},{I_{DB}}} \right\}}\left\{ {\exp \left( { - \frac{{{x^{{\alpha _1}}}T\left( {\beta {I_{{r_1}}} + {I_{{r_2}}} + {\sigma ^2}} \right)}}{{\beta {P_1}}}} \right)} \right\} \\
 &\mathop  = \limits^{\left( b \right)} \exp \left( { - \frac{{{x^{{\alpha _1}}}T{\sigma ^2}}}{{\beta {P_1}}}} \right){{\cal L}_{{I_{BB}}}}\left( {\frac{{{x^{{\alpha _1}}}T}}{{{P_1}}}} \right){{\cal L}_{{I_{DB}}}}\left( {\frac{{{x^{{\alpha _1}}}T}}{{\beta {P_1}}}} \right),
\end{aligned}
\end{equation}
where (a) follows that ${h_{{x}}}\sim\exp{(1)}$ and (b) is because of the independence of interference $I_{r_k}$. Likewise, we can obtain the CDF of the SINR for other three cases, as given by
\begin{equation}\label{Apx3:eq3}
\begin{aligned}
 &\mathbb{P} \left[ {SIN{R_{\bar B}}\left( x \right) \ge T} \right] \\
 &= \exp \left( { - \frac{{{x^{{\alpha _1}}}T{\sigma ^2}}}{{{P_1}}}} \right) \hspace{-1mm} {{\cal L}_{{I_{\bar B\bar B}}}} \hspace{-1mm} \left( {\frac{{{x^{{\alpha _1}}}T}}{{{P_1}}}} \right) \hspace{-1mm} {{\cal L}_{{I_{\bar D\bar B}}}} \hspace{-1mm} \left( {\frac{{{x^{{\alpha _1}}}T}}{{{P_1}}}} \right),
\end{aligned}
\end{equation}
\begin{equation}\label{Apx3:eq4}
\begin{aligned}
 &\mathbb{P} \left[ {SIN{R_D}\left( x \right) \ge T} \right] \\
 &= \exp \left( { - \frac{{{x^{{\alpha _2}}}T{\sigma ^2}}}{{{P_2}}}} \right) \hspace{-1mm} {{\cal L}_{{I_{BD}}}} \hspace{-1mm} \left( {\frac{{{x^{{\alpha _2}}}T\beta }}{{{P_2}}}} \right) \hspace{-1mm} {{\cal L}_{{I_{DD}}}} \hspace{-1mm} \left( {\frac{{{x^{{\alpha _2}}}T}}{{{P_2}}}} \right),
\end{aligned}
\end{equation}
\begin{equation}\label{Apx3:eq5}
\begin{aligned}
 &\mathbb{P} \left[ {SIN{R_{\bar D}}\left( x \right) \ge T} \right] \\
 &= \exp \left( { - \frac{{{x^{{\alpha _2}}}T{\sigma ^2}}}{{{P_2}}}} \right) \hspace{-1mm} {{\cal L}_{{I_{\bar B\bar D}}}} \hspace{-1mm} \left( {\frac{{{x^{{\alpha _2}}}T}}{{{P_2}}}} \right) \hspace{-1mm} {{\cal L}_{{I_{\bar D\bar D}}}} \hspace{-1mm} \left( {\frac{{{x^{{\alpha _2}}}T}}{{{P_2}}}} \right).
\end{aligned}
\end{equation}

With the consideration of the Laplace transform of the interference, ${{\cal L}_{{I_{BB}}}}\left( s \right)$ can be calculated, as formulated by
\begin{equation}\label{Apx3:eq6}
\begin{aligned}
 {{\cal L}_{{I_{BB}}}}\left( s \right) & = \mathbb{E}_{{I_{BB}}}\left\{ {\exp \left( { - s{I_{BB}}} \right)} \right\}  \\
 &= \mathbb{E}_{\left\{ {{\Phi _1},{h_i}} \right\}}\left\{ {\exp \left( { - s \hspace{-2mm} \sum\limits_{i \in {\Phi _1}\backslash {b_0}} \hspace{-2mm} {{P_1}{h_i}r_i^{ - {\alpha _1}}} } \right)} \right\} \\
 &=\mathbb{E}_{{\Phi _1}}\left\{ {\prod\limits_{i \in {\Phi _1}\backslash {b_0}} {{\mathbb{E}_{{h_i}}}} \left( {\exp \left( { - s{P_1}{h_i}r_i^{ - {\alpha _1}}} \right)} \right)} \right\} \\
 &\mathop  = \limits^{\left( c \right)} \mathbb{E}_{{\Phi _1}}\left\{ {\prod\limits_{i \in {\Phi _1}\backslash {b_0}} {\frac{1}{{1 + s{P_1}r_i^{ - {\alpha _1}}}}} } \right\},
\end{aligned}
\end{equation}
where (c) follows that ${h_{{i}}}\sim\exp{(1)}$. With the help of the probability generating function~\cite{ref22}, the following relationship can be attained, as expressed by
\begin{equation}\label{Apx3:eq7}
 \mathbb{E} \left\{ {\prod\limits_\Phi  {f\left( x \right)} } \right\} = \exp \left( { - \lambda \int_{{\mathbb{R}^2}} {\left( {1 - f\left( x \right)} \right)} dx} \right).
\end{equation}

Then substituting $s = \frac{{{x^{{\alpha _1}}}T}}{{{P_1}}}$ into \eqref{Apx3:eq6} and using \eqref{Apx3:eq7}, Eq.~\eqref{Apx3:eq6} can be reformulated as
\begin{equation}\label{Apx3:eq8}
 {{\cal L}_{{I_{BB}}}} \hspace{-1mm} \left( \hspace{-0.5mm} {\frac{{{x^{{\alpha _1}}}T}}{{{P_1}}}} \hspace{-0.5mm} \right) \hspace{-1mm} = \hspace{-1mm} \exp \hspace{-1mm} \left( \hspace{-0.5mm} { - 2\pi {\lambda _1} \hspace{-2mm} \int_{{x}}^\infty \hspace{-2mm} {\left( \hspace{-0.5mm} {\frac{1}{{1 \hspace{-1mm} + \hspace{-1mm} {T^{ - 1}}{{\left( {\frac{y}{x}} \right)}^{{\alpha _1}}}}}} \hspace{-0.5mm} \right)} ydy} \hspace{-0.5mm} \right).
\end{equation}

Through integral transformation $v = {\left( {{x^{{\alpha _1}}}T} \right)^{\frac{{ - 2}}{{{\alpha _1}}}}}{y^2}$, Eq.~\eqref{Apx3:eq8} can be simplified, as presented by
\begin{equation}\label{Apx3:eq9}
\begin{aligned}
 {{\cal L}_{{I_{BB}}}} \hspace{-1mm} \left( \hspace{-0.5mm} {\frac{{{x^{{\alpha _1}}}T}}{{{P_1}}}} \hspace{-0.5mm} \right) \hspace{-1mm} &= \hspace{-1mm} \exp \hspace{-1mm} \left( \hspace{-1mm} { - \pi {x^2}{\lambda _1}{{T}^{\frac{2}{{{\alpha _1}}}}} \hspace{-2mm} \int_{{{T}^{\frac{{ - 2}}{{{\alpha _1}}}}}}^\infty \hspace{-1mm} {\left( \hspace{-0.5mm} {\frac{1}{{1 + {v^{\frac{{{\alpha _1}}}{2}}}}}} \hspace{-0.5mm} \right)} dv} \right) \\
 \hspace{-1mm} &= \hspace{-1mm} \exp \left( { - \pi {x^2}{\lambda _1} Q \left( {T,{\alpha _1}} \right)} \right).
\end{aligned}
\end{equation}
Similarly, we can have
\begin{equation}\label{Apx3:eq10}
\begin{aligned}
&{{\cal L}_{{I_{DB}}}}\left( {\frac{{{x^{{\alpha _1}}}T}}{{\beta {P_1}}}} \right) = \\
&\exp \left( \hspace{-0.5mm} { - \pi {x^{\frac{{2{\alpha _1}}}{{{\alpha _2}}}}}{\lambda _2}{{\left( {\frac{{T{P_2}}}{{\beta {P_1}}}} \right)}^{\frac{2}{{{\alpha _2}}}}} \hspace{-2mm} \int_{{{\left( {\frac{{\beta {B_1}}}{T}} \right)}^{\frac{2}{{{\alpha _2}}}}}}^\infty \hspace{-2mm}  {\left( {\frac{1}{{1 + {v^{\frac{{{\alpha _2}}}{2}}}}}} \right)} dv} \hspace{-0.5mm} \right).
\end{aligned}
\end{equation}

Combining \eqref{Apx3:eq1} and \eqref{eq6:1}, the SINR coverage ${S_{B}}\left( T \right)$ can be formulated as \eqref{eq8}. With the same method, the SINR coverage in other cases, i.e., ${S_{\bar B}}\left( T \right)$, ${S_{D}}\left( T \right)$, and ${S_{\bar D}}\left( T \right)$ can be obtained as \eqref{eq9}--\eqref{eq11} respectively, based on which {Theorem}~\ref{Lem:2} has been proved rigidly.
\end{IEEEproof}

\end{document}